\documentclass[aps,prl,twocolumn,showpacs,amsmath,amssymb,superscriptaddress]{revtex4-1}

\usepackage{graphicx}
\usepackage{color}
\usepackage[normalem]{ulem}

\graphicspath{{.}{./EPS/}}

\begin{document}

\title{Motion of an Impurity in a Bose-Einstein Condensate with Weyl Spin-Orbit Coupling:
{Non-collinear} Drag Force and Anisotropic Critical Velocity}

\author{Renyuan Liao}
\affiliation{Fujian Provincial Key Laboratory for Quantum Manipulation and New Energy Materials,
College of Physics and Energy, Fujian Normal University, Fuzhou 350108, China}

\author{Oleksandr Fialko}
\affiliation{Dodd-Walls Centre for Photonics and Quantum Technology, Institute of Natural and Mathematical Sciences and Centre for Theoretical Chemistry and
Physics, Massey University, Auckland 0745, New Zealand}

\author{Joachim Brand}
\affiliation{Dodd-Walls Centre for Photonics and Quantum Technology, New Zealand Institute for
Advanced Study, and Centre for Theoretical Chemistry and Physics, Massey University, Auckland
0745, New Zealand}

\author{Ulrich Z\"ulicke}
\affiliation{School of Chemical and Physical Sciences and Dodd-Walls Centre for Photonics and
Quantum Technology, Victoria University of Wellington, PO Box 600, Wellington 6140, New Zealand}

\date{\today}

\begin{abstract}
We {consider} the motion of a point-like impurity through a three-dimensional
two-component Bose-Einstein condensate subject to Weyl spin-orbit coupling. {Using
linear-response theory, we} calculate the drag force felt by the impurity and the associated
anisotropic critical velocity  from the spectrum of elementary
excitations. The drag force is shown to be generally not collinear with the velocity of the impurity.
This {unusual behavior} is {a consequence of}
condensation {into a finite-momentum state} due to the spin-orbit coupling.
\end{abstract}


\maketitle

Degenerate quantum gases of neutral atoms~\cite{DAL99,BLO08}, polaritons~\cite{CAR13} as well
as the recently discovered condensation of light~\cite{CAR13} have provided new opportunities for
studying superfluidity.  One of the most remarkable manifestations of superfluidity is that impurities
immersed into such systems propagate without dissipation if their velocities do not exceed the
Landau critical velocity~\cite{Lan47}
  \begin{eqnarray}
     v_c={\rm min}_q\left[\frac{\omega(q)}{q}\right],
     \label{Eq:Landau}
  \end{eqnarray}
where $\omega(q)$ is the spectrum of elementary excitations. As long as the impurity moves slower
than the critical velocity, the superfluid cannot absorb any of its energy and therefore the impurity
motion is frictionless. Experiments with atomic Bose-Einstein condensates (BEC) have provided
evidence for a critical velocity associated with emission of elementary excitations~\cite{RAM99,
ONO00,DRI10,DES12,JEN14} as well as more complex  excitations like vortices and
solitons~\cite{INO01,ENG07,NEE10,RAM11,WRI13}. Intense theoretical efforts~\cite{FRI92,KAG00,
PIT04,ROB05,CUC06,CAR06,SAC06,AND09,TEM09,RYA10,SCH12,GOR12,CHE12,KAM13} have
been undertaken to study the stability of superfluidity and explain the mechanisms of dissipation in
BECs.

Recently, the experimental realization of various synthetic gauge fields including one-dimensional
and two-dimensional  spin-orbit coupling (SOC) in quantum gases~\cite{DAL11,VIC13,GOL14,
HUA15} enabled the prediction of a number of novel interesting properties  in these new types of
condensates~\cite{GOL14,ZHA15}. Among them is condensation of Bose atoms at some finite
momentum, thus breaking simultaneously the conventional U(1) gauge symmetry associated with
condensation as well as rotational symmetry. Moreover, SOC breaks the Galilean invariance of the
system~\cite{BIA12,KHA14,ZHA15}, making the applicability of the Landau criteria of superfluidity in
the new BECs questionable. This calls for a better understanding of the critical velocity and
dissipation mechanism of this new superfluid.

In this Letter, we examine superfluidity in a two-component Bose gas with three-dimensional Weyl
SOC  by studying the drag force felt by a moving point-like impurity~\cite{LEG06}. The Weyl SOC
can be realized using powerful quantum technology~\cite{GOL14}. Here we calculate the drag force
using linear response theory from the elementary excitation spectrum through the dynamical
structure factor~\cite{PIT04}. The drag force demonstrates the presence of an anisotropic critical
velocity. We also find that the drag force is not generally collinear with the velocity of the impurity, in
stark contrast to a conventional superfluid. This fact can be used to probe SOC by the scattering of
heavy molecules by the condensate.

The Weyl-type SOC takes its name from a seminal study by Hermann Weyl~\cite{WEY29} predicting
fermions with a high degree of symmetry. Although there is currently no evidence for Weyl fermions
to exist as fundamental particles in our universe, Weyl-like quasiparticles have been detected
recently in condensed-matter systems~\cite{XU15,LV15}. In light of these discoveries, the study of
Weyl SOC in ultra-cold atom systems becomes particularly relevant, since the ability to manipulate
the Weyl-SOC strength creates interesting opportunities for the discovery of effects not predicted in
the realm of particle physics.

The second-quantized Hamiltonian for the BEC with a point-like impurity moving with velocity
$v$ is
\begin{eqnarray}
H&=&\int d^3\mathbf{r}\Psi^\dagger\left[\left(-\frac{\hbar^2\nabla^2}{2m}-\mu\right)I+\lambda
\vec{\sigma}\cdot\mathbf{P}\right]\Psi\nonumber\\
 &+&\int d^3\mathbf{r}\left[\frac{g}{2}n^2+(g_{\uparrow\downarrow}-g)n_\uparrow n_\downarrow
 +g_i n \delta({\bf r}-{\bf v}t)\right] .
\label{Eq:model}
\end{eqnarray}
Here $\Psi(\mathbf{r})=(\psi_\uparrow,\psi_\downarrow)^T$ is the two-component condensate
quantum field, $I$ is the $2\times 2$ identity matrix, $n({\bf r})=n_{\uparrow}({\bf r})+n_{\downarrow}
({\bf r})\equiv \psi^{\dagger}_\uparrow\psi_\uparrow+\psi^{\dagger}_\downarrow\psi_\downarrow$ is
the density operator, $\mu$ is the chemical potential, $\lambda$ is the strength of the spin-orbit
coupling, $g_i$ is the particle-impurity coupling constant, and the strengths of the intra-species
interaction and inter-species interaction are $g$ and $g_{\uparrow\downarrow}$, respectively. For
brevity, we set $\hbar=2m=1$ from now on.

The time-dependent mean-field Gross-Pitaevskii (GP) equation found from Eq.~(\ref{Eq:model})
reads
\begin{eqnarray}
&&\left[(-i\partial_t-\nabla^2 - \mu)I-i\lambda\vec{\sigma}\cdot{\vec{\nabla}}\right]\Psi_0+gn_0 \Psi_0
\nonumber\\&+&(g_{\uparrow\downarrow}-g) \begin{pmatrix}n_{0\downarrow}&0\\0&n_{0\uparrow}
\end{pmatrix}\Psi_0+g_i\delta({\bf r}-{\bf v}t)\Psi_0=0 .
\end{eqnarray}
To proceed, we split the field $\Psi_0({\bf r},t)=\Phi_0({\bf r})+\Phi({\bf r},t)$, where $\Phi_0({\bf r})=
\sqrt{\frac{n_0}{2}}(1,1)^Te^{i\mathbf{K}\cdot\mathbf{r}}$ is the mean-field solution without the
impurity, and $\Phi(\mathbf{r},t)$ is the perturbation caused by the impurity. Without loss of
generality, we choose the condensation momentum to be $\mathbf{K}=(-\lambda/2,0,0)$, and the
chemical potential becomes $\mu=n_0(g+g_{\uparrow\downarrow})/2-K^2$. Linearizing GP in
$\Phi$, we obtain
\begin{eqnarray}
&&\left[\left(-i\partial_t-\nabla^2+K^2+\frac{gn_0}{2}\right)I-i\lambda\vec{\sigma}\cdot\vec{\nabla}+
\frac{g_{\uparrow\downarrow}n_0}{2}\sigma_x\right]\Phi\nonumber\\
&+&\left(\frac{g_{\uparrow\downarrow}n_0}{2}\sigma_x+\frac{gn_0}{2}I\right)e^{2i\mathbf{K}\cdot
\mathbf{r}}\Phi^*+g_i\delta({\bf r}-{\bf v}t)\Phi_0=0.
\end{eqnarray}
The ansatz $\Phi({\bf r},t)=e^{i\mathbf{K}\cdot\mathbf{r}}\sum_\mathbf{q}\varphi_\mathbf{q}e^{i
\mathbf{q}\cdot(\mathbf{r}-{\bf v}t)}$ yields
\begin{eqnarray}\label{eq:linGPtrans}
&&\left[(-\mathbf{q}\cdot\mathbf{v}+q^2+2K^2+2\mathbf{K}\cdot\mathbf{q})I+\lambda\vec{\sigma}
\cdot(\mathbf{K+q})\right]\varphi_\mathbf{q}\nonumber\\
&&+\left(\frac{gn_0}{2}I+\frac{g_{\uparrow\downarrow}n_0}{2}\sigma_x\right)(\varphi_\mathbf{q}+
\varphi_{\mathbf{-q}}^*)=-g_i\sqrt{\frac{n_0}{2}}\begin{pmatrix}1\\1\end{pmatrix}.
\end{eqnarray}
Combining Eq.~(\ref{eq:linGPtrans}) with its complex conjugate, we obtain
\begin{equation}\label{Eq:f_q}
\left[ {\mathcal H}_\mathbf{q} - \mathbf{q}\cdot\mathbf{v} \, {\mathcal I} \right] \left( \begin{array}{c}
\varphi_{\mathbf{q}\uparrow} \\\varphi_{\mathbf{q}\downarrow} \\ \varphi_{-\mathbf{q}\uparrow}^*\\\varphi_{-\mathbf{q}\downarrow}^* \end{array} \right) =-g_i\sqrt{\frac{n_0}{2}} \begin{pmatrix} 1 \\ 1 \\1\\1\end{pmatrix}
\quad ,
\end{equation}
with the matrices
\begin{equation}
{\mathcal H}_\mathbf{q} = \begin{pmatrix} M_\mathbf{q} & B\\ B & M_{-\mathbf{q}}^* \end{pmatrix}
\quad , \quad {\mathcal I} = \left(\begin{array}{cc} I & 0 \\ 0 & -I \end{array} \right) \,\, .
\end{equation}
Here $B=\frac{gn_0}{2}I+\frac{g_{\uparrow\downarrow}n_0}{2}\sigma_x$ and
\begin{equation}
   M_\mathbf{q}=(q^2+2K^2+2\mathbf{K}\cdot\mathbf{q})I
   +\lambda\vec{\sigma}\cdot(\mathbf{K+q})+B \,\, .
\end{equation}
Solving Eq.~(\ref{Eq:f_q}) for $\varphi_{\bf q \uparrow}$ and $\varphi_{\bf q \downarrow}$ , the force acting on the impurity is found as
\begin{eqnarray}
\mathbf{F}&=&-\int d^3\mathbf{r}\Psi_0^\dagger \vec{\nabla}\left[g_i\delta(\mathbf{r}-\mathbf{v}t)
\right]\Psi_0 \equiv g_i\nabla|\Psi_0(\mathbf{r},t)|^2_{\mathbf{r}=\mathbf{v}t}\nonumber \,\, , \\ \\
&\approx&g_i\sqrt{\frac{n_0}{2}}\sum_\mathbf{q}i\mathbf{q}\left(\varphi_{\mathbf{q}\uparrow}+
\varphi_{\mathbf{q}\downarrow}+\varphi_{\mathbf{-q}\uparrow}^*+\varphi_{\mathbf{-q}\downarrow}^*
\right) \,\, , \nonumber \\ \label{eq:LRdrag}
&=&-\frac{g_i^2n_0}{2}\sum_\mathbf{q}i\mathbf{q}\sum_{ij} \left( \left[ {\mathcal H}_\mathbf{q} -
(\mathbf{q}\cdot\mathbf{v} +i0^+)\, {\mathcal I} \right]^{-1}\right)_{ij} \,\, . \nonumber \\
\end{eqnarray}
The infinitesimal imaginary part was added following the usual causality rule~\cite{PIT04,CAR06,
PEI14}.

According to the fluctuation-dissipation theorem, the drag force should be related to the fluctuation
properties (i.e., the spectrum of elementary excitations) of the unperturbed system. This fact will
assist us to analyze the drag force in more detail. We consider the system without the impurity by
setting $g_i=0$ in Eq.~(\ref{Eq:model}). The partition function of the system can be conveniently
casted as imaginary time field integral~\cite{SIM06} $\mathbf{Z}=\int d[\psi_\sigma^*,\psi_\sigma]
e^{-S[\psi_\sigma^*,\psi_\sigma]}$ with the action given by $S=\int_0^\beta d\tau\left[\int d\mathbf{r}
\sum_\sigma\psi_\sigma^*\partial_\tau\psi_\sigma+H(\psi_\sigma^*,\psi_\sigma)\right]$, where
$\tau=it$ is the imaginary time. We replace the Bose field with a static part and a fluctuating part as
$\psi_\sigma=\phi_{\sigma0}+\phi_\sigma$~\cite{LIA13,LIA14}. Within the Bogoliubov approximation,
we expand the action up to quadratic orders in the fluctuating fields, and approximate the action as
$S\approx S_{\rm eff}=S_0+S_g$. Here $S_0$ is the saddle point action containing only the static
fields $\phi_{\sigma0}$, and $S_g$ is the Gaussian action containing fluctuating fields $\phi_\sigma$
of quadratic orders. By defining column vectors $\Xi_q=(\phi_{\mathbf{K+q}\uparrow},\phi_{\mathbf{K
+q}\downarrow},\phi_{\mathbf{K-q}\uparrow}^*,\phi_{\mathbf{K-q}\downarrow}^*)^T$, we may write
the Gaussian action in a compact form as $S_g=\frac{1}{2}\sum_{\mathbf{q},iw_n}\Xi_q^\dagger
\mathcal{G}^{-1}\Xi_q$, where $q=(\mathbf{q},iw_n)$ with $w_n=2n\pi/\beta$ being bosonic
Matsubara frequencies. The inverse Green's function is given by $\mathcal{G}^{-1}(\mathbf{q},iw_n)
= \mathcal{H}_\mathbf{q} - i \omega_n \, \mathcal{I}$. Comparison with Eq.~(\ref{eq:LRdrag}) yields
the drag force in terms of the Green's function of the unperturbed system:
\begin{eqnarray}
\mathbf{F}=-\frac{g_i^2n_0}{2}\sum_\mathbf{q}i\mathbf{q}\sum_{ij}\mathcal{G}_{ij}(\mathbf{q},iw_n
\rightarrow \mathbf{q}\cdot\mathbf{v}+i0^+).
\end{eqnarray}

Within Bogoliubov theory, the dynamical structure factor of the unperturbed system can
be evaluated as~\cite{LIA13}
\begin{eqnarray}
S(\mathbf{q},iw_n)&=&N^{-1} \langle\delta \rho(\mathbf{q},iw_n)^\dagger\delta\rho(\mathbf{q},iw_n)
\rangle_0\nonumber\\
&=&\sum_{i,j}{\cal G}_{ij}(\mathbf{q},iw_n).
\end{eqnarray}
Thus the drag force can be also expressed in terms of the dynamic structure factor:
\begin{eqnarray}
\mathbf{F}=-\frac{g_i^2n_0}{2}\sum_\mathbf{q}i\mathbf{q} \, S(\mathbf{q},iw_n\rightarrow \mathbf{q}
\cdot\mathbf{v}+i0^+).
\end{eqnarray}
The spectrum $\omega_i({\bf q})$ of elementary excitations is found by solving ${\rm Det}
[\mathcal{G}^{-1} (\mathbf{q},iw_n)]=0$ with subsequent analytic continuation $i\omega_n\rightarrow
\omega_i({\bf q})$. Using this fact, we arrive at the following expression for the drag force:
\begin{eqnarray}
\mathbf{F}=2\pi g_i^2n_0\sum_\mathbf{q}\mathbf{q}\sum_{i=1}^4\frac{J(\mathbf{q},\omega_i({\bf q}))
\delta(\mathbf{q}\cdot\mathbf{v}-\omega_i({\bf q}))}{\prod_{j\neq i}\left[\omega_i({\bf q})-\omega_j({\bf
q})\right]} .
\label{thirteen}
\end{eqnarray}
Here we used the abbreviation
\begin{eqnarray}
J(\mathbf{q},iw_n)&=&q^2(iw_n+2\lambda q_x)^2-(q^2+\lambda^2)(q^4+\lambda^2q_x^2)
\nonumber\\
& &-(g-g_{\uparrow\downarrow})n_0(q^4+\lambda^2q^2-\lambda^2q_y^2) \,\, .
\label{Eq:J}
\end{eqnarray}

Let us first calculate the drag force in the absence of SOC by setting $\lambda=0$. In this case,
the four branches of excitations are the Bogoliubov-type modes $\omega^0_{1,2}=\pm q \sqrt{q^2+(g
+g_{\uparrow\downarrow}) n_0}$ and $\omega^0_{3,4}=\pm q\sqrt{q^2+(g-g_{\uparrow\downarrow})
n_0}$. The former is the spectrum of density waves propagating with the speed of sound $c=\sqrt{(g
+g_{\uparrow\downarrow})n_0}$, while the latter is the spectrum of spin waves. Using these
analytical expressions, we evaluate the sum in Eq.~(\ref{thirteen}) and find the drag force
$\mathbf{F}_0={\bf v}g_i^2n_0v(1-c^2/v^2)^2/(16\pi){\Theta}(v-c)$, in agreement with
Ref.~\cite{PIT04}. The drag force in this case is collinear with the impurity's direction of motion.
Note also that the lower-lying spin-wave mode is not excited because the impurity couples only to
density waves.

The presence of SOC modifies the above result. Due to condensation into a finite-momentum state,
the ground state breaks rotational symmetry, and the spectrum of elementary excitations becomes
anisotropic. For our choice of condensate momentum, the spectrum is invariant under flipping the
direction of $q_y$ and/or $q_z$, namely $\omega_i(q_x,\pm q_y,\pm q_z)=\omega_i(q_x,q_y,q_z)$.
As a result, the $x$-direction is distinguished from the $y$ and $z$ axes. To be specific, let's assume
that the impurity moves along the $z$ axis. We can write $\delta(\mathbf{q\cdot v}-\omega_i)=\delta
(q-q_{z0})/|v_z-\partial \omega_i/\partial q_z|$, where $q_{z0}=h_i(q_x,q_y^2)$ is some function
reflecting symmetry properties, and the detailed form of $h_i$ is unimportant for our further analysis.
Carrying out the integration in  Eq.~(\ref{thirteen}), one immediately finds that $F_y$ vanishes and
both $F_x$ and $F_z$ survive, by considering the integration of odd or even functions within a
symmetrical interval. This argument can be repeated for different directions of the velocity yielding
an additional contribution to the drag force along the $x$ axis.  Therefore, in addition to the
conventional force component along the velocity vector, a force component along $x$ axis is
generated, due to the asymmetrical excitation spectrum with regard to $x$ axis. For small spin-orbit
coupling, corrections in the drag force brought about by SOC may be estimated by performing an
expansion in the parameter $\lambda$. Up to the first non-vanishing order in the SOC strength, after
lengthly but straightforward calculations, we obtain ${\bf F}={\bf F}_0+{\bf F}_{\parallel}+{\bf F}_x$
with ${\bf F}_{\parallel}\approx \hat{v}\mathcal{O}(\lambda^2)$ and ${\bf F}_x \hat{x}\mathcal{O}
(\lambda^3)$. Evidently, SOC  produces an additional drag along the direction of condensation.

\begin{figure}[t]
\includegraphics[width=1.0\columnwidth]{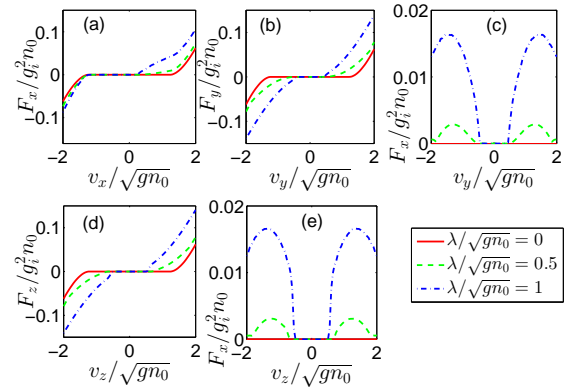}
\caption{(color online) Drag force (measured in units of $g_i^2n_0$) experienced by an impurity moving in a Weyl-spin-orbit-coupled
BEC with condensate wave vector parallel to the negative $x$ direction and $g_{\uparrow
\downarrow}/g=1/2$. Results for different spin-orbit coupling strength $\lambda$ are given. For the case of the impurity velocity being parallel to the $x$ axis, only
the $x$ component $F_x$ of the drag force is finite even for $\lambda\ne 0$ [panel (a)]. When the
velocity is along  the $y$ axis, the drag force has components along the $y$ axis [panel (b)] and, for
finite $\lambda$, also along the $x$ axis [panel(c)]. When the velocity is along the $z$ axis, the drag
force has finite components in the $z$ and $x$ directions when $\lambda \ne 0$ [panels (d) and
(e)].}
\label{fig1}
\end{figure}

\begin{figure}[b]
\includegraphics[width=1.0\columnwidth]{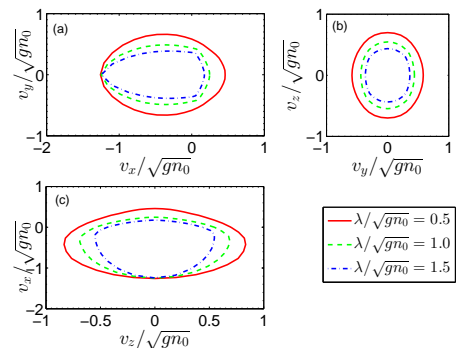}
\caption{(color online) Directional dependence of the critical velocity for an impurity moving in a
Weyl-spin-orbit-coupled BEC with condensate wave vector parallel to the negative $x$ direction
and $g_{\uparrow\downarrow}/g=1/2$. }
\label{fig2}
\end{figure}

To substantiate the above argument we now calculate the drag force (\ref{thirteen}) numerically. For
our numerics we choose $gn_0$ as the energy scale and $\sqrt{gn_0}$ as the momentum scale.
Let us first consider
simple situations where the velocity of the moving impurity is along the $x$, $y$ and $z$ axes,
respectively. Results obtained for these cases are shown in Fig.~\ref{fig1}. When the velocity is
along the $x$ axis, the drag force is also along $x$ axis, and is asymmetrical between positive and
negative direction of velocity, reflecting the spontaneously broken symmetry of the ground state with
finite condensate momentum in the negative $x$ direction. It requires more force to drag the impurity
against the direction of condensation than along with it. When the velocity is along the $y$ axis or
the $z$ axis, the magnitude of force does not change upon reversing the direction of the velocity.
However, it is remarkable that in both cases a non-vanishing force component along the $x$ axis
emerges. Fig.~\ref{fig1} also shows that there exists a critical velocity $v_c$ below which there is no
drag force. Its magnitude  decreases when the strength of SOC is increased. We calculated the
critical velocity in three orthogonal planes by examining the lower bound of the velocity where the
drag force becomes nonzero. These results are presented in Fig.~$\ref{fig2}$. As shown in panel
(a), the critical velocity along the \emph{negative} $x$ axis remains unchanged as SOC strength is
increased, however, it decreases significantly along the \emph{positive} $x$ axis. As can be seen in
panel (b), the critical velocity in the $y z$ plane is slightly deformed from a circle, signalling the
inequivalence between $y$ and $z$ directions.

\begin{figure}[b]
\includegraphics[width=1.0\columnwidth]{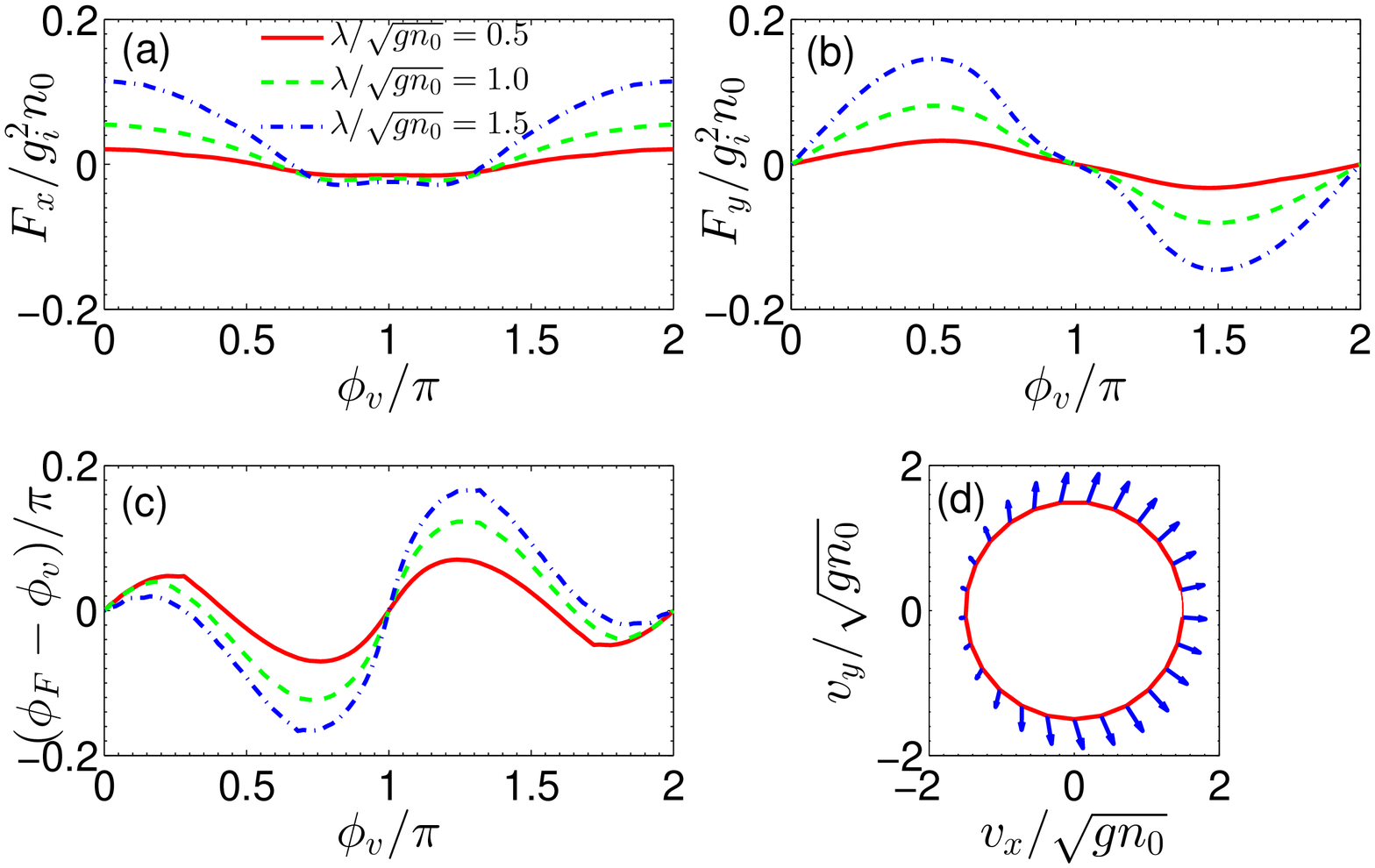}
\caption{(color online) Cartesian components of the drag force for an impurity moving in the $xy$
plane at azimuthal angle $\phi_v$ with speed $v = 1.5\,\sqrt{gn_0}$: (a) $F_x$; (b) $F_y$. (c)
Difference between the azimuthal angles of drag force and velocity; $\phi_F-\phi_v$. (d)
Visualization of the drag force for $\lambda=1.5 \,\sqrt{gn_0}$. Blue arrows indicate the force
vectors for velocities corresponding to points on the red circle. Here $g_{\uparrow\downarrow}/
g=1/2$ was assumed.}
\label{fig3}
\end{figure}

Let us now examine the drag force in more detail. Fixing the velocity of the impurity to lie in the
$xy$ plane we show the behavior of the corresponding drag force in Fig.~\ref{fig3}. Here we
define the azimuth of the drag force to be $\phi_F=arg(F_x+iF_y)$, i.e.~the angle in the $xy$ plane, and the azimuth of the velocity to be
$\phi_v=arg(v_x+iv_y)$. In panel (a), the $x$ component for the scaled drag force $F_x$ has the
symmetry of $F_x(\pi-\phi_v)=F_x(\pi+\phi_v)$, namely it has reflection symmetry with respect to the
$x$ axis. The $y$ component $F_y$ entails the symmetry of $F_y(\pi-\phi_v)=-F_y(\pi+\phi_v)$, as
indicated in panel (b). The $z$ component of the drag force vanishes. In panel (c), we show the
difference between the azimuth of the drag force and the azimuth of the velocity. It is quite remarkable
that the direction of the drag force is not aligned with the velocity, as is the case in a conventional
superfluid. For a better visualization, we show the force vector in panel
(d), where the arrow sitting on constant circle of speed indicates the force vector.

\begin{figure}[t]
\includegraphics[width=1.0\columnwidth]{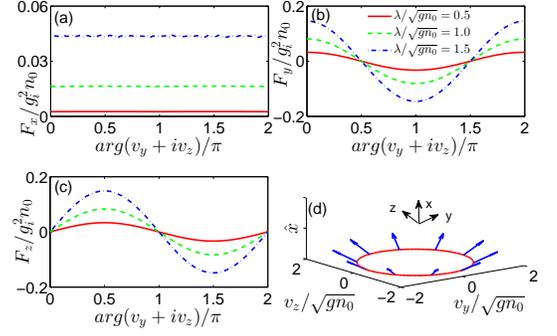}
\caption{(color online) Cartesian components of the drag force for an impurity moving in the $yz$
plane with speed $v = 1.5\,\sqrt{gn_0}$: (a) $F_x$; (b) $F_y$; (c) $F_z$. (d) Visualization of the force
vector for $\lambda=1.5\, \sqrt{gn_0}$. The force vector for different velocity directions is indicated
by arrows sitting on the circle of constant speed. Here $g_{\uparrow\downarrow} /g=1/2$ was
assumed.}
\label{fig4}
\end{figure}

Now we fix the velocity vector to lie in the $yz$ plane. The drag force is shown in Fig.~\ref{fig4}.
Panel (a) illustrates that the drag force has an $x$ component that is independent of the direction of
the velocity within the $yz$ plane for a fixed small spin-orbit coupling strength $\lambda$. For
large $\lambda$, $F_x$ oscillates slightly with varying directions of the velocity in $yz$ plane. This
inequivalence between $y$ axis and $z$ axis is due to the breaking of spin-rotational invariance by
the non-linear interaction potential $g-g_{\uparrow\downarrow}\ne 0$. In panel (b), $F_y$ is shown to
be symmetric  with respect to $y$ axis while antisymmetric with respect to $z$ axis. In panel (c),
$F_z$ is antisymmetric with respect to $y$ axis and symmetric with respect to the $z$ axis.
Interestingly, as can been seen in panel (d), the direction of the force is not lying in the $yz$ plane,
but is tilted towards the $x$ axis.

\begin{figure}[b]
\includegraphics[width=1.0\columnwidth]{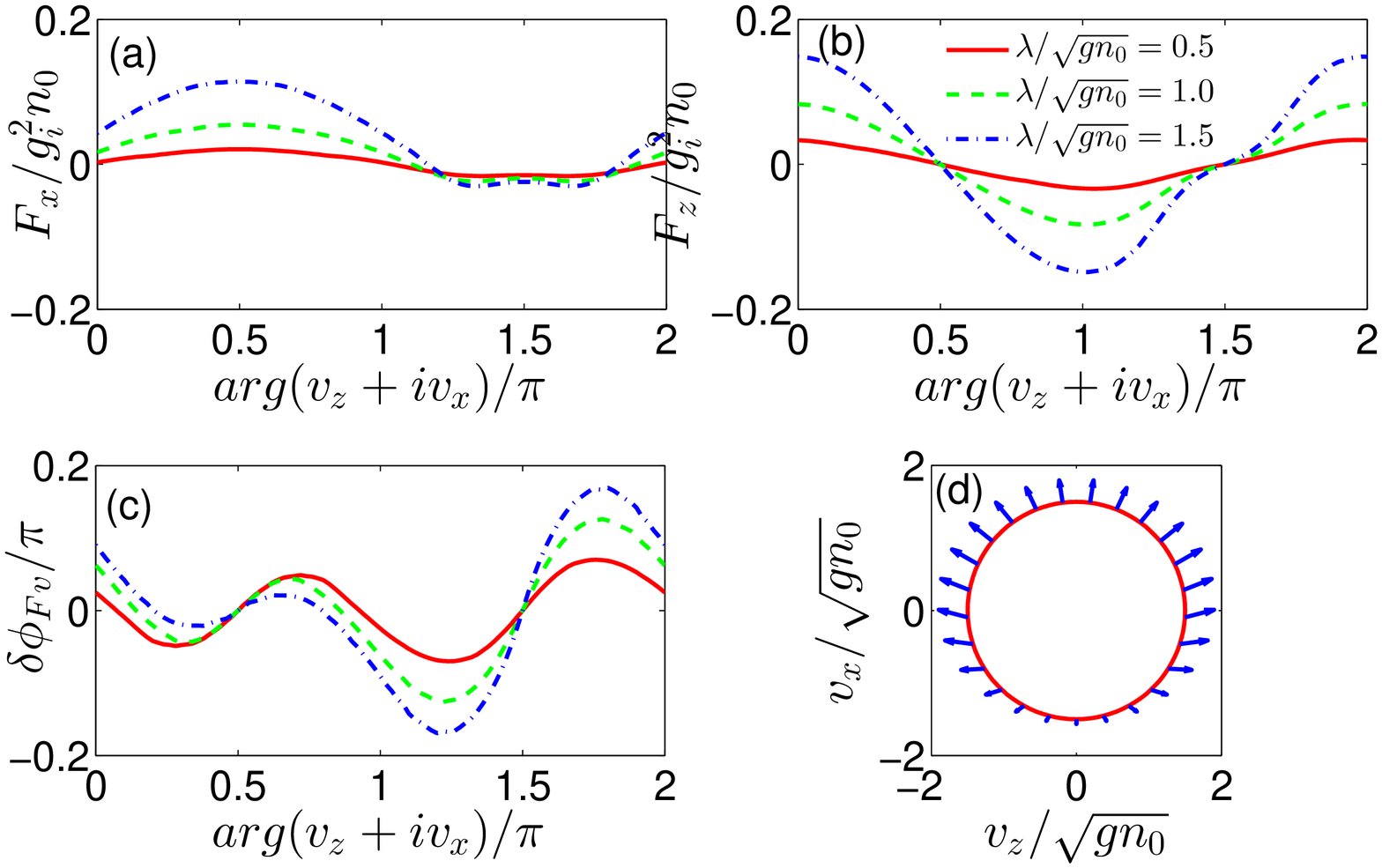}
\caption{(color online) Cartesian components of the drag force for an impurity moving in the $zx$
plane with speed $v = 1.5\,\sqrt{gn_0}$: (a) $F_x$; (b) $F_z$. (c) Difference between the azimuthal
angles of the drag force and the velocity; $\delta \phi_{Fv}\equiv arg(F_z+iF_x)-arg(v_z+iv_x)$. (d)
Visualization of the force vector for $\lambda=1.5\,\sqrt{gn_0}$, with force vectors indicated by
arrows sitting on the circle of constant speed in the $zx$ plane. Here $g_{\uparrow\downarrow}/
g=1/2$ was assumed.}
\label{fig5}
\end{figure}

Results for the case when the velocity vector lies in the $zx$ plane are shown in
Fig.~\ref{fig5}. The behavior of the drag force looks quite similar to the situation when the
velocity is in the $xy$ plane. Panel (a) illustrates that $F_x$ is symmetric with respect to $x$ axis,
while panel (b) shows that $F_z$ is antisymmetric with respect to the $z$ axis. As seen in
panel (c), the difference of azimuthal angles for the force and velocity is antisymmetric with respect
to the $z$ axis. In panel (d), the force vector is visualized.

In summary, we have studied the motion of a point-like impurity in a three-dimensional
two-component BEC with Weyl SOC. We calculated the drag force and the associated critical
velocity. At small SOC strength, we showed that the drag force can be decomposed into two parts.
One is along the direction of the moving velocity, and the other one is along the direction of the
condensation momentum. Hence, unlike in non-spin-orbit-coupled superfluids, the drag force is not
generally collinear with the velocity of the impurity. This unusual feature can be utilized to probe
SOC in bosonic superfluids.

R.~L. acknowledges funding from the NSFC under Grants No.\ 11274064 and NCET-13-0734.
O.~F. was supported by the Marsden Fund (contract MAU1205), administered by the Royal Society
of New Zealand.

%

\end{document}